\documentclass[showpacs,prl,superscriptaddress,twocolumn]{revtex4}
\usepackage{graphicx,amsmath}

\begin{document}

\title{Diamagnetic orbital response of mesoscopic silver rings}

\author{R. Deblock} \affiliation{Laboratoire de Physique des Solides, Associ\'e
au CNRS, B\^atiment 510, Universit\'e Paris-Sud, 91405 Orsay, France}
\author{R. Bel} \affiliation{Laboratoire de Physique des Solides, Associ\'e au
CNRS, B\^atiment 510, Universit\'e Paris-Sud, 91405 Orsay, France} \author{B.
Reulet} \affiliation{Laboratoire de Physique des Solides, Associ\'e au CNRS,
B\^atiment 510, Universit\'e Paris-Sud, 91405 Orsay, France} \author{H.
Bouchiat} \affiliation{Laboratoire de Physique des Solides, Associ\'e au CNRS,
B\^atiment 510, Universit\'e Paris-Sud, 91405 Orsay, France} \author{D. Mailly}
\affiliation{CNRS Laboratoire de Photonique et Nanostructures, 196 avenue
Ravera, 92220 Bagneux, France}

\pacs{73.20.Fz, 73.23.Ra}

\begin{abstract} We report measurements of the flux-dependent orbital magnetic
susceptibility of an ensemble of $10^5$ disconnected silver rings at 217 MHz.
Because of the strong spin-orbit scattering rate in silver this experiment is a
test of existing theories on ensemble averaged persistent currents. Below 100 mK the rings exhibit
a magnetic signal with a flux periodicity of $h/2 e$ consistent with  averaged
persistent currents, whose amplitude is estimated to be of the order of 0.3 nA.
The sign of the oscillations indicates unambiguously diamagnetism in the vicinity of zero
magnetic field. This sign is  a priori not consistent with theoretical predictions for
average persistent currents. We discuss several  possible  explanations of this result.
  \end{abstract}

\maketitle

At low temperature, electrons in metallic mesoscopic samples keep their phase
coherence on a length $L_{\Phi}$ which is larger than the sample size. The transport and thermodynamic properties of the system are then sensitive to
interference between electronic wave functions and present spectacular
signatures of this phase coherence. To measure these effects the ring geometry
is particularly suitable. Indeed in the presence of a magnetic flux $\Phi$
through the ring the periodic boundary conditions for electronic wave functions
acquire a phase factor $2 \pi \Phi/\Phi_0$ with $\Phi_0=h/e$ the flux quantum
\cite{intro}. As a consequence  the free energy $F$ of the system is flux
dependent which leads to the  existence of a non-dissipative current
$I=-\partial F/\partial \Phi$, the  persistent current, which is a periodic
function of $\Phi$ \cite{buttiker83} and can be detected by  measuring the induced magnetic moment. 

Persistent currents in mesoscopic rings have been studied for over 10 years theoretically \cite{cheung89,montambaux90,ambegaokar90,schmid91, vonoppen91,altshuler91, kravtsov00} and experimentally addressing the question of either the typical current (measured on a single or a small number of rings) \cite{chandrasekhar91,mailly93,jariwala01,rabaud01},  or of the ensemble averaged  persistent current (measured on arrays of rings) \cite{levy90, reulet95,jariwala01,deblock01}. However theory and experiment  still do not agree. One important subject of disagreement is the sign of the average value of these persistent currents. Recent experiments on GaAs rings \cite{deblock01} indicate  a sign corresponding to low field diamagnetism. In contrast  theoretical calculations which include repulsive field interactions predict paramagnetism  \cite{ambegaokar90,schmid91,eckern95}.
A recent theory \cite{kravtsov00} shows that additional currents may be generated in rings due to the rectification of a high frequency non equilibrium noise. These currents,(as also suggested in reference \cite{mohanty99}) are not experimentally distinguishable from persistent currents. This
mechanism yields a diamagnetic sign and gives the right amplitude for the average current measured in GaAs rings. An important 
feature  is that the sign of this 
current should change from diamagnetic to paramagnetic depending on the strength of spin-orbit
scattering.  In contrast, the thermodynamic equilibrium
persistent currents are insensitive to spin-orbit scattering. This motivates the comparison between the orbital magnetism of mesoscopic rings fabricated  in different  materials having various strength of spin-orbit
scattering. The experiments on GaAs rings dealt with the case of weak spin-orbit interactions. Previous experiments on Au rings \cite{jariwala01} addressed the case of strong spin-orbit but were  however not conclusive concerning this sign because the current was not averaged on a sufficient  number of rings. 
The present experiment was designed to resolve this question: using Ag (a material with strong spin-orbit interactions) and a large number of rings.

\begin{figure}[tb]
	\begin{center} 	
		\includegraphics[width=5cm]{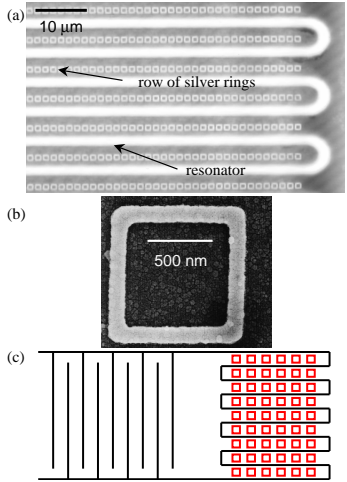}
	\end{center}
	\caption{(a) Photograph of a part of the resonator with the silver rings. (b) 
	Image obtained by scanning electron microscopy of one silver ring. (c) Schematic picture 
	of the resonator with the square rings on the inductive part. The resonator
has an 
	inductive part (meander line) and a capacitive part (comb-like structure).}
	\label{fig:anneauagresbis} 
\end{figure}

We have investigated the magnetic response at low temperature of a collection
of silver rings, with spin-orbit scattering length  much shorter than their circumference  in contrast with the
GaAs case. The magnetic response of the rings is detected by a resonant method
: the rings are coupled to the inductive part of a superconducting
microresonator made of niobium deposited on sapphire. This resonator has a
resonance frequency of 217 MHz with a quality factor of $2 \, 10^5$ below 1 K.
A schematic picture of the resonator, composed of an inductance (meander line)
and a capacitance (comb-like structure), with the rings on the inductive part
is shown on Fig. \ref{fig:anneauagresbis} (c). The detailed fabrication and
characterization of this type of resonator has been described elsewhere
\cite{deblock01}. 

By measuring the shift of the resonance
frequency $f$ of the resonator induced by the presence of the rings, we have access to $\chi$ the averaged
magnetic susceptibility of the silver rings, according to:
\begin{equation} \label{eq:df}
	\frac{\delta f}{f}=-\frac{1}{2} N k_m \chi \end{equation}  where $N$ is the
number of rings coupled to the resonator, $k_m$ is a coefficient characterizing
the coupling of one ring to the inductive part of the resonator  \cite{deblock01}.
This quantity is recorded as
a function of the amplitude of a DC magnetic field produced by a
superconducting coil. The sample is placed in a dilution refrigerator with a
base temperature of 30 mK.  
\begin{table*}[tb]
	\begin{center}
		\begin{tabular}{|c|c|c|c|c|c|c|c|}
		\hline
		perimeter $L$ & width $w$ & thickness $h$ & $l_e$ & $D=v_F l_e/3$ & $E_c=h
D/L^2$ & 
		$L_{so}$ & $L_{\Phi}(T=40 mK)$\\
		\hline
		4 $\mu$m & 130 nm & 70 nm & 40 $\pm$ 5 nm & 0.018 m$^2$.s$^{-1}$ & 54
mK & 575 nm & 
		13 $\mu$m \\	
		\hline
		\end{tabular}
	\caption{Characteristics of the silver rings. The mean free path $l_e$ is
deduced from the resistance of silver wires fabricated together with the rings and with the same width. The spin-orbit length $L_{so}$ and the phase coherence length $L_{\Phi}$ are extracted from  previous weak-localisation measurements on silver wires also fabricated in the same evaporator and using  the same bulk silver \cite{gougam00}.}
	\label{tab:anneauag}
	\end{center} 
\end{table*}

The fabrication  involved two steps of lithography. First the
superconducting resonator is fabricated by optical lithography and tested at
low temperature. PMMA is then  spin-coated on top of the resonator. A high
resolution e-beam writer is used to pattern rows of  rings aligned
between the meanders of the inductive part of the resonator. After metal
deposition and lift-off \cite{spec} we obtain around $1.5 \, 10^5$ silver rings
aligned with the inductance of the resonator,  on the same substrate. With this configuration the quality factor of the bare resonator is
preserved in contrast with previously measured GaAs/AlGaAs rings
\cite{deblock01}. Moreover thanks to the alignment of the rings with
the inductance the coupling coefficient $k_m$ is the same for all the rings. As
a consequence the rings experience the same AC field, another improvement over the experiments on GaAs rings. Images of the resonator
and the rings are shown on Fig. \ref{fig:anneauagresbis}. During the lift-off
process some of the rings are cut or damaged so that one can estimate the
number of rings producing a signal to be between $10^5$ and $1.4 \, 10^5$. This
very large number of rings insures that  the measurement is dominated
by the ensemble average of the magnetic susceptibility of the rings ( which was
not the case on previous experiments on an array of 30 Au rings \cite{jariwala01}). 
 The different characteristic lengths  of the silver rings are given in table\ref{tab:anneauag}, they verify $L_{\phi} >L > L_{SO}$.  This experiment is  in to the strong spin-orbit scattering regime   at magnetic field below 100 gauss. In this low field regime the magnetic length defined as: $L_H^2=\Phi_0/H $ is larger than the spin-orbit scattering length. Data were taken within this field range during slow DC magnetic field sweeps (0.02
G/s). We moreover restricted ourselves to magnetic field less than 30
G in order to prevent vortex trapping in the superconducting resonator which  give rise to inhomogeneities of DC magnetic field  on the different rings. In addition all
data was acquired while decreasing the  absolute value of the magnetic field to
minimize hysteresis effects.  We also had  to operate the resonator in the over-coupled regime with respect to the RF generator and strongly attenuate the RF power injected  to
less than 0.1 pW. This was done  to prevent electron heating by the RF radiation and minimize  spurious signal  originating from  defects in the meander line.
\begin{figure}[bt]
	\begin{center} 		
		\includegraphics[width=7.5cm]{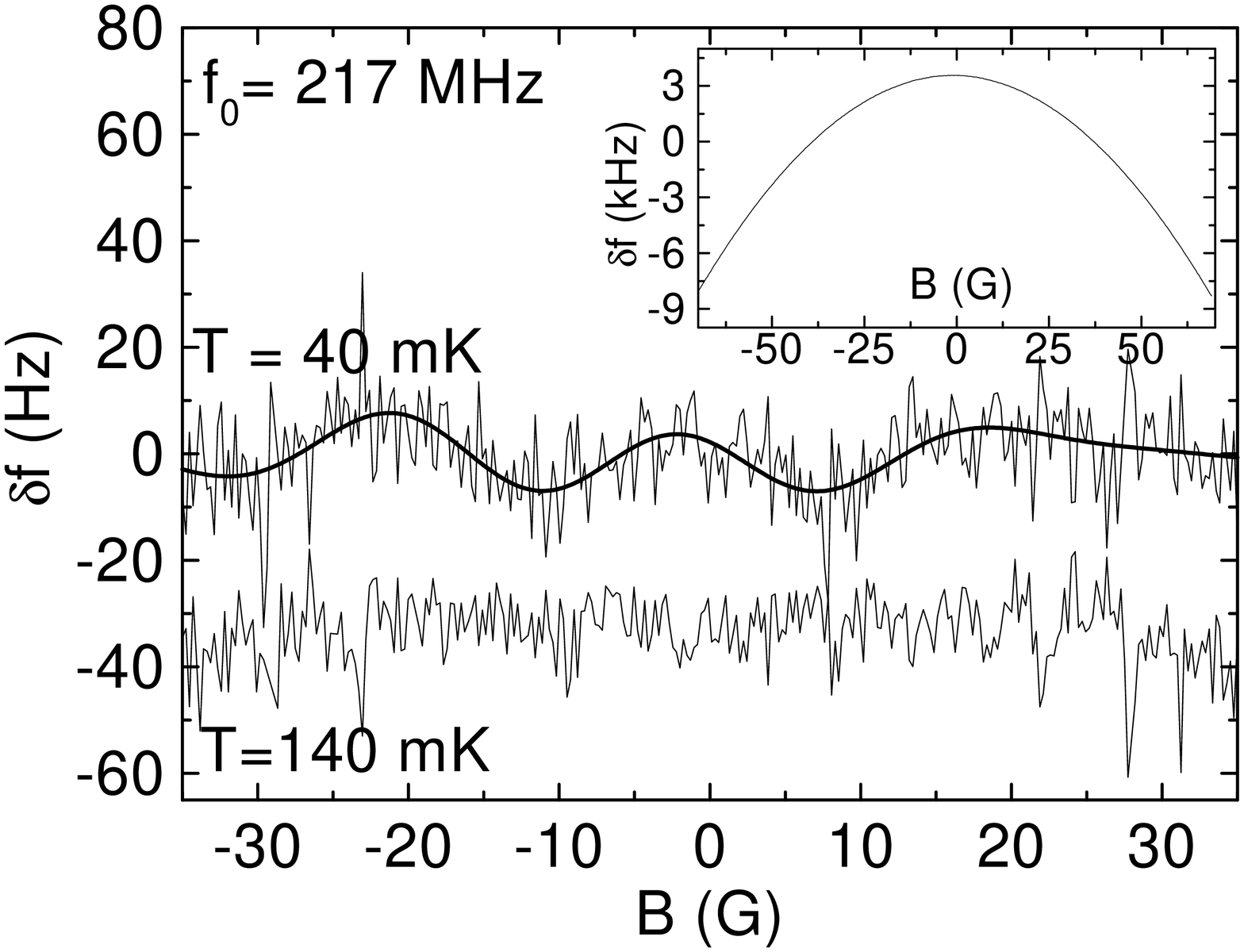}
	\end{center}
	\caption{Resonance frequency shift due to the silver rings at 40 mK and 140 mK 
	versus magnetic field. This data is obtained after subtraction of the parabolic signal 
	characteristic of the bare resonator shown in the inset and is obtained after averaging 50 curves. 
	The smooth curve on the low temperature data is
	obtained by numerical filtering. Curves are shifted for clarity. Inset : Frequency shift 
	of the resonator with the ring. The parabolic behaviour is due to the flux dependence 
	of the diamagnetic response of the bare resonator.}
	\label{fig:ringag} 
\end{figure}
\begin{figure}[tb]
	\begin{center} 		
		\includegraphics[width=7.5cm]{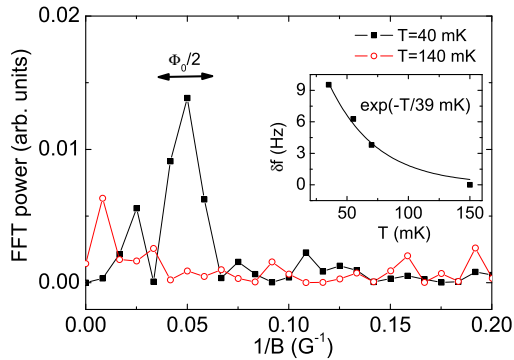}
	\end{center}
	\caption{Fourier transform of the frequency shift due to the rings at 40 mK
and 140 mK. Inset: temperature dependance of the frequency shift due to the rings. The data are consistent with an exponential decay with a temperature scale of 40 mK.}
	\label{fig:cpringagfft} 
\end{figure}

\begin{figure}[tb]
	\begin{center} 		
		\includegraphics[width=7.5cm]{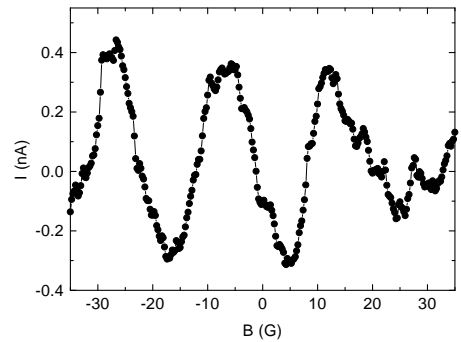}
	\end{center}
	\caption{Average persistent currents through the rings reconstituted from the field dependence of the resonance frequency  in Fig.2  according to expression  \ref{eqi} after high pass filtering at 0.025 $Gauss^{-1}$ and integration of the signal.}
	\label{fig4} 
\end{figure}

The resonance frequency shift versus magnetic field has two parts : the first
one is due to the magnetic field dependence of the magnetic penetration length in the resonator. This part 
has been checked to be the same with or without the rings. In the small
magnetic field range explored here, it is well approximated by a parabola (see
inset of Fig. \ref{fig:ringag} and Ref. \cite{deblock01}). Periodic oscillations are superimposed on top of this signal.  To focus on
these oscillations the contribution of the bare resonator is subtracted.  After
averaging 50 curves, we get the data of Fig. \ref{fig:ringag}. At low
temperature ($T = 40$ mK) the oscillations are visible. Note that the ``noise''
on the curve is reproducible : it corresponds to  imperfections of the
resonator which form Josephson junctions sensitive to magnetic field. It
indicates that we are close to the limit of detection of our experimental
setup. At higher temperature ($T = 140$ mK) the oscillations are strongly
reduced as also visible  on the Fourier transform of the two
previous curves (Fig. \ref{fig:cpringagfft}). At low temperature the
periodicity measured is consistent with half a flux quantum in the area
enclosed in a ring, as expected for the average susceptibility 
\cite{montambaux90,kamenev94}.
From the value of the maximum frequency shift $\delta_{\Phi} f =
f(\Phi_0/4)-f(\Phi=0)$ and given the coupling coefficient $k_m$
\cite{deblock01} we deduce the amplitude of the variation with magnetic field
of the magnetic susceptibility $\delta_{\Phi} \chi(\omega)= 5.3 \pm 0.9 \, 
10^{-24}$ m$^3$. In order to compare this value to existing theories on orbital
magnetism we make the assumption that the signal measured is due to persistent
currents:
\begin{equation}
\frac{\partial I (\Phi)}{\partial \Phi}= -\frac{\delta f (\Phi)}{2f} \frac{\mathcal{L}}{ N
\mathcal{M}^2} \label{eqi} \end{equation} 
with $\mathcal{L}=0.05 \mu$H the estimated
inductance of the resonator and $\mathcal{M}=0.14$ pH the calculated mutual
inductance between one ring and the inductive part of the resonator. It is then possible to deduce the flux dependence of the  average persistent current depicted in Fig.\ref{fig4} which oscillates with a periodicity $\Phi_0/2$ and an amplitude  $|I_0|=0.33 \pm 0.05$ nA.   The current is  \textit{diamagnetic} at
low field.  Note that the sign of the susceptibility  oscillations can be
determined in a completely unambiguous way in this experiment,   since it is
directly related to the sign of the resonance frequency shift measured. Another check comes
 from the  field dependence of the diamagnetic  signal of the bare resonator (see inset of Fig.\ref{fig:ringag}). The temperature dependence of the oscillations is shown on  inset of Fig.
\ref{fig:cpringagfft}. The signal cannot be detected above 100mK. From the small number of data points, we can only say that this  
temperature decrease of the signal is consistent with an exponential behavior with a temperature
scale of the order of 40 mK, which is close to the value of the Thouless energy
$E_c = h D/L^2$ where $D$ is the diffusion coefficient. 

We now compare our result with theoretical predictions. Due to the very small
value of the mean level spacing $\Delta$ in metal rings, theoretical
predictions based on non-interacting electrons \cite{altshuler91}, which lead
to a current of amplitude $\Delta/\Phi_0$, are not able to explain the value of
the current measured. In the case of interacting electrons \cite{ambegaokar90},
the expected value of the persistent current is of the order of $0.2
E_c/\Phi_0=0.04$ nA, which is within an order of magnitude of the
experimental value. However theory predicts in this case a paramagnetic current
for repulsive interactions, with or without strong spin-orbit coupling
\cite{ambegaokar90,schmid91}.  Considering instead attractive interactions, superconducting
fluctuations in silver with a critical temperature of the order of 50 nK, would
lead to a diamagnetic current (therefore in agreement with our experiment) with the same amplitude as the repulsive case. 
Moreover it has been recently  suggested \cite{imry02} that superconducting fluctuations could give rise to a much larger current than the initial predictions of reference  \cite{ambegaokar90}. If one now considers the possibility of non equilibrium noise induced currents, according to ref. \cite{kravtsov00}, it is clear that this mechanism cannot explain the present experimental results since a paramagnetic average current is predicted 
in the strong spin-orbit limit. 

All the previous analyses are valid in the limit of zero frequency. Our
experiment however is performed at the resonance frequency of the resonator,
i.e. 217 MHz. This frequency is larger than $1/\tau_{\Phi}$ and not negligible
compared to $1/\tau_D$ ($1/\tau_{\Phi}=100$ MHz and $1/\tau_D=1$ GHz). In this
frequency range it is reasonable to expect extra contributions to the magnetic
response of Aharonov-Bohm rings. This point is already contained in the early work of
Altshuler, Aronov and Spivak describing the $\Phi_0/2$ oscillations in the
average conductance $\Delta G(\omega)$ of a ring of perimeter $L=2 \pi R$
excited by an electromotive force \cite{altshuler81,aronov87}: \begin{equation}
	\Delta G(\omega)=-\frac{e^2}{\pi \hbar} \frac{L_{\Phi}(\omega)}{L}
\sum_{l=-\infty}^{+\infty}
	\frac{1}{\pi} \frac{R/L_{\Phi}(\omega)}{\left(
\frac{R}{L_{\Phi}(\omega)}\right)^2
	+\left( l+\frac{2 \Phi}{\Phi_0} \right)^2} \end{equation} with
$L_{\Phi}(\omega)^2=L_{\Phi}^2/(1+i \omega \tau_{\Phi})$. At finite frequency
the  imaginary part $G^{''}(\omega)$of $\Delta G(\omega)$ is non zero and one can associate to this imaginary conductance a non
dissipative current which response  function reads : \begin{equation}
	\frac{\partial I}{\partial \Phi} = \omega G^{''}(\omega) \end{equation} The
first harmonics is of the order of $E_c/\Phi_0$ and changes sign with
spin-orbit.  More precisely for parameters close to the experimental ones  $\omega \tau_{\Phi}=2$ and $L_{\Phi}/R=10$, one finds numerically
a diamagnetic current of amplitude 0.1 $E_c/\Phi_0=0.02nA$ of the order of the thermodynamic current. Note that
weak-localisation experiments on silver wires at finite frequency are in
agreement with this analysis \cite{pieper92}. 
However it was pointed out by Efetov \textit{et al.} that this result is  in principle only
valid for connected geometries \cite{efetov}.  Finally we would like to point out that the contribution of
electron-electron interactions, which plays an essential role in the 
thermodynamic current, has not been examined at finite frequency. 

To conclude we have demonstrated in this experiment the existence of $\Phi_0/2$ periodic orbital 
magnetism in silver rings at 217 MHz and low temperature. The signal is
consistent  with \textit{diamagnetic} averaged persistent currents which
does not  agree  with theoretical predictions in the presence of repulsive interactions.
 This sign can be explained either by the existence of very weak superconducting fluctuations  or by the value of the  measurement frequency which is
larger than the inverse phase coherence time.
\begin{acknowledgments} We acknowledge fruitful 
discussions with S. Gu\'eron, V.E. Kravtsov and G. Montambaux. We would like to
thank the quantronic group at SPEC (Saclay) for letting us use their
nanolithography facility.  \end{acknowledgments}


\begin{thebibliography}{0}
\bibitem{spec} The deposition of silver and the lift-off process was done  with
the facility of the quantronic group at SPEC in Saclay. \cite{gougam00}.
\end{thebibliography}
\end{document}